\documentclass[reprint,superscriptaddress,nofootinbib,amsmath,amssymb,aps,pra,longbibliography,tightenlnes]{revtex4-2}
\usepackage{graphicx}
\usepackage{dcolumn}
\usepackage{bm}
\usepackage{bbm}
\usepackage{amsmath}
\usepackage{amssymb}
\usepackage{amsfonts}
\usepackage{latexsym}
\usepackage{color}
\usepackage[caption=false]{subfig}
\usepackage[colorlinks=true,linkcolor=blue,citecolor=blue,urlcolor=blue]{hyperref}
\usepackage{braket}
\usepackage{amsmath}
\usepackage[T1]{fontenc}
\usepackage[utf8]{inputenc}

\begin{document}
\preprint{APS/123-QED}

\title
{Loophole-free Bell-inequality violation between atomic states in cavity-QED systems mediated by hybrid atom-light entanglement}

\author{Pei-Zhe Li}
\email{p.li@oist.jp}
\affiliation{Okinawa Institute of Science and Technology Graduate University, 1919-1 Tancha, Onna-son, Okinawa 904-0495, Japan}

\author{Soumyakanti Bose}
\affiliation{NextQuantum Innovation Research Center, Department of Physics \& Astronomy, Seoul National University, Gwanak-ro 1, Gwanak-gu, Seoul 08826, Korea}
\affiliation{Department of Physics, SRM University, Amaravati, Andhra Pradesh - 522540, India}

\author{Hyunseok Jeong}
\affiliation{NextQuantum Innovation Research Center, Department of Physics \& Astronomy, Seoul National University, Gwanak-ro 1, Gwanak-gu, Seoul 08826, Korea}

\author{William J. Munro}
\affiliation{Okinawa Institute of Science and Technology Graduate University, 1919-1 Tancha, Onna-son, Okinawa 904-0495, Japan}

\author{Kae Nemoto}
\affiliation{Okinawa Institute of Science and Technology Graduate University, 1919-1 Tancha, Onna-son, Okinawa 904-0495, Japan}
\affiliation{National Institute of Informatics, 2-1-2 Hitotsubashi, Chiyoda-ku, Tokyo 101-8430, Japan}

\author{Nicol\`o Lo Piparo}
\email{nicolo.lopiparo@oist.jp}
\affiliation{Okinawa Institute of Science and Technology Graduate University, 1919-1 Tancha, Onna-son, Okinawa 904-0495, Japan}

\date{\today}

\begin{abstract}

We present a feasible and scalable approach to testing Bell nonlocality and implementing device-independent quantum key distribution (DI-QKD) between distant atomic states in cavity-based architectures, mediated by hybrid atom-light entanglement.
We develop a full theoretical model that incorporates realistic sources of noise—such as transmission loss, limited light–matter coupling efficiency, and imperfect detection.
Our analysis shows that strong Bell-Clauser–Horne–Shimony–Holt (CHSH) violations and secure key generation over tens of kilometers are within reach using current or near-term technology.
These results position cavity-based platforms with coherent-state encodings as a promising foundation for future scalable, DI quantum communication networks.
\end{abstract}

\keywords{}

\maketitle

\section{\label{intro}Introduction}

Quantum technologies have advanced rapidly in recent years, enabling the precise control of individual photons, atoms, and solid-state qubits \cite{Zhong2020, Madsen2022, Yin2017, Debnath2016, Bernien2017, Browaeys2020, Arute2019, Veldhorst2015, Sukachev2017}.  
Beyond their practical applications, these systems also provide powerful testbeds for exploring fundamental aspects of quantum mechanics, including the limits of locality and realism highlighted by Bell's theorem \cite{Einstein1935,Bell1964}.  
A violation of Bell inequalities not only challenges classical descriptions of nature but also provides a certified source of quantum correlations that can be harnessed for secure communication.  
In particular, device-independent quantum key distribution (DI-QKD) exploits Bell nonlocality as a resource: a sufficiently strong Bell violation guarantees secrecy without requiring trust in the internal workings of the devices \cite{Mayers1998, Barrett2005, Acin2006, Acin2006a, Acin2007, Pironio2009, Masanes2011, Vazirani2014, Miller2016}. 

However, current Bell-test implementations achieve only modest violations and operate over limited distances, which severely constrains their applicability to DI-QKD \cite{Hensen2015, Giustina2015, Rosenfeld2017, Storz2023,Zapatero2023}.  
Realizing scalable and practically useful DI-QKD therefore requires new methods to distribute entanglement robustly in the presence of losses and detector imperfections.  
Although several DI-QKD experiments have recently been reported \cite{Nadlinger2022,Zhang2022,Liu2022}, their performance remains far from that required for practical deployment.  
In particular, Nadlinger \textit{et al.}~\cite{Nadlinger2022}, Zhang \textit{et al.}~\cite{Zhang2022}, and Liu \textit{et al.}~\cite{Liu2022} achieved secret key rates (SKRs) of $\sim3.3$ bits/s over 3.5~m of fiber, $\sim8.5\times10^{-4}$~bits/s over 700~m of fiber, $\sim2.6$~bits/s over 220~m, and , respectively.  
These implementations rely on single-photon sources, for which long-distance transmission and high collection efficiency remain major experimental challenges. 
Consequently, exploring alternative, more loss-tolerant approaches to entanglement distribution is essential for advancing both foundational quantum tests and practical quantum communication. 

A promising direction lies in continuous-variable (CV) schemes \cite{Glancy2004, Loock2006, Dias2017, Furrer2018, Seshadreesan2020}.  
Recently, entangled atom–light cat states have been experimentally demonstrated \cite{Hacker2019}, offering a route to robust entanglement distribution through CV transmission channels \cite{Bose2024, Bera2025}.  
Moreover, these techniques naturally extend to rotation-symmetric bosonic codes (RSBCs) \cite{Grimsmo2020, Li2024}, which provide enhanced protection against photon loss—the dominant error source in optical fibers.

In this work, we propose and analyze a Bell test protocol based on coherent states within a cavity-QED architecture.  
While the protocol can be extended to various RSBCs, we focus on the experimentally realized one-loss cat codes that can be generated deterministically in cavity-QED systems \cite{Hacker2019}.  
To demonstrate the practicality and robustness of our scheme, we incorporate several realistic imperfections into our theoretical analysis, including photon loss during light–matter interactions, fiber transmission, and detection.  
We evaluate its performance by calculating the Clauser–Horne–Shimony–Holt (CHSH) parameter $S$ for cat codes of different loss orders under realistic conditions, and we further estimate the achievable DI-QKD key rates.

The paper is organized as follows:  In Section~\ref{mc}, we describe the protocol for generating Bell states over long distances.  Section~\ref{sec:bell} reviews Bell nonlocality and derives the corresponding form for our scheme. 
In Section~\ref{per}, we present numerical results for the CHSH parameter and DI-QKD key rates under realistic conditions.  Finally, Section~\ref{con} summarizes our conclusions and discusses potential extensions.

\section{\label{mc}Protocol}

Typically, Bell tests rely on pairs of entangled systems that are spatially separated.  The separation distance between the remote parties must be sufficiently large to ensure a space-like separation, thereby closing the locality loophole \cite{Brunner2014}. In practice, reproducing such conditions in a laboratory requires a platform capable of generating and manipulating entanglement between well-defined, spatially distinct quantum systems.  Here, we describe a platform that fulfills these requirements, based on cavity-QED and RSBCs.

The protocol exploits a specific light–matter interaction achievable in the cavity-QED regime \cite{Duan2004,Reiserer2015,Hacker2019}.  The system consists of a single atom or ion in an optical cavity, modeled as a three-level system with two ground states, $\ket{\downarrow}$ and $\ket{\uparrow}$, and one excited state, $\ket{e}$.  The state $\ket{\uparrow}$ is resonant and strongly coupled to the bare cavity mode, whereas $\ket{\downarrow}$ is off-resonant and effectively decoupled due to the ground-state splitting.

Under these conditions, a coherent state $\ket{\alpha}$, with the same frequency as the bare cavity mode, is reflected from the cavity.  The parameter $\alpha$ represents the field amplitude, and $|\alpha|^2$ corresponds to the mean photon number.  Upon reflection, the coherent state acquires a state-dependent phase shift, determined by whether the atom (or ion) is in $\ket{\uparrow}$ or $\ket{\downarrow}$. 
The interaction can be expressed as \cite{Duan2004,Reiserer2015,Hacker2019}:
\begin{eqnarray}  \label{eq1}
    \begin{aligned}
        &\ket{\uparrow}\ket{\alpha} \rightarrow \ket{\uparrow}\ket{\alpha}, \;\;\;\;\;\;\;\;\;
        \ket{\downarrow}\ket{\alpha} \rightarrow \ket{\downarrow}\ket{-\alpha}.
    \end{aligned}
\end{eqnarray}
This operation corresponds to a CNOT-like gate, where the atomic state acts as the control qubit and the photonic mode as the target. 

For an atom (ion) prepared in the superposition $(\ket{\uparrow}+\ket{\downarrow})/\sqrt{2}$, the reflection process generates the hybrid atom–light entangled state $(\ket{\uparrow}\ket{\alpha}+\ket{\downarrow}\ket{-\alpha})/\sqrt{2}$.  Here, we encode logical states using coherent states with opposite amplitudes: $\ket{\bar{0}}=\ket{\alpha}$ and $\ket{\bar{1}}=\ket{-\alpha}$.  By repeating this procedure, one can create multi-component cat states, which offer enhanced protection against photon-loss errors \cite{Li2023a}. The cat codes generated via this method can be written as:
\begin{eqnarray} \label{eq11}
    \begin{aligned}
        &\left|\left.\bar{0}_{2^m}\right\rangle\right.=\frac{1}{\sqrt{N_{2^m}}}\sum_{k=0}^{2^m-1}{\ket{e^{i\frac{2k\pi}{2^m}}\alpha}},\\
        &\left|\left.\bar{1}_{2^m}\right\rangle\right.=\frac{1}{\sqrt{N_{2^m}}}\sum_{k=0}^{2^m-1}{\ket{e^{i\frac{(2k+1)\pi}{2^m}}\alpha}},
    \end{aligned}
\end{eqnarray}
where $l=2^m-1$ defines the loss order of the code, representing the number of photon losses that can be corrected.  For example, the code with logical states $\ket{\bar{0}}=\ket{\alpha}$ and $\ket{\bar{1}}=\ket{-\alpha}$ corresponds to the 0-loss cat code.  According to \cite{Li2023a}, cat codes with $l=2^m-1$ can be generated via $m+1$ rounds of light–matter interaction.

\begin{figure}[htb]
\includegraphics[width=0.48\textwidth]{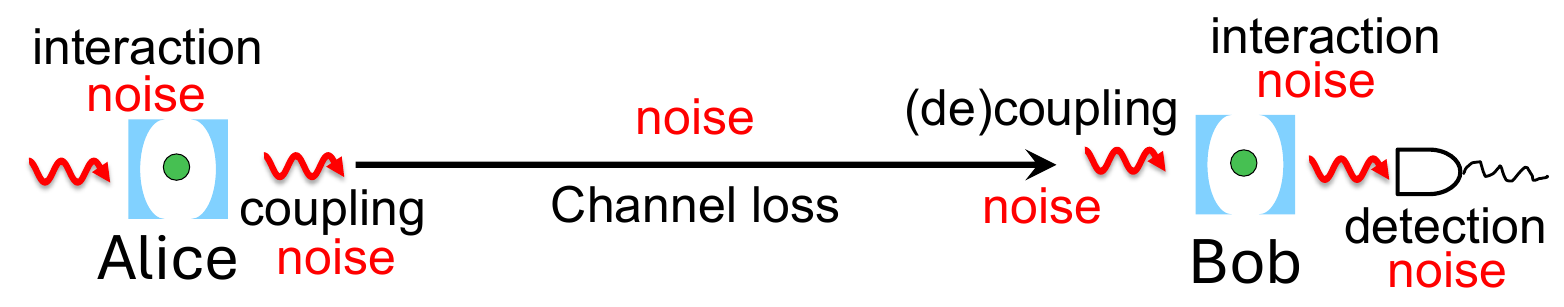}
 \caption{\label{figbp} Schematic of the protocol for generating entangled states between distant nodes using cavity-QED. Potential sources of error are indicated at the corresponding stages of the process. }
\end{figure}

In our protocol depicted in Fig. \ref{figbp}, two distant parties, Alice and Bob, separated by a distance $L$, each possess a trapped atom or ion inside an optical cavity.  Initially, both atoms are prepared in the equal superposition state $(\ket{\uparrow}+\ket{\downarrow})/\sqrt{2}$.

Alice first reflects a coherent state $\ket{\alpha}$ from her cavity.  Through the light–matter interaction described above, this process entangles the reflected optical mode with Alice's atomic qubit.  The resulting optical field is then transmitted through an optical fiber to Bob.

Upon arrival, Bob performs a syndrome measurement by reflecting the incoming light from his cavity $m$ times, which allows detection of photon losses up to the loss order $l$.  After the syndrome measurement, the light mode is reflected once more from another cavity on Bob's side, undergoing the same conditional interaction.  This final step entangles the traveling light mode with Bob's atomic qubit, thereby completing the atom–light–atom entanglement process.  The quantum state after this interaction can be expressed as:
\begin{eqnarray} \label{eq2}
    (\ket{\uparrow\uparrow}_{AB}+\ket{\downarrow\downarrow}_{AB})\ket{\alpha}+(\ket{\uparrow\downarrow}_{AB}+\ket{\downarrow\uparrow}_{AB})\ket{-\alpha}.
\end{eqnarray}

Finally, Bob performs a logical-Z measurement on the light modes to distinguish the codewords.  This logical-Z measurement can be implemented using unambiguous state discrimination (USD) \cite{Li2023a}.  After USD, Alice and Bob share an entangled state that depends on the outcome of the state discrimination.  To preserve coherence during transmission, Alice requires a quantum memory to store her atomic state until the optical mode reaches Bob and all subsequent operations are completed.  The Bell test can then be performed on the distributed remote entangled pairs. 

In this work, the Bell test is conducted on the atomic entangled states rather than the light–matter entangled states. This choice is motivated by the fact that atomic states can generally be measured with very high efficiency and fidelity in arbitrary bases, whereas performing measurements on CV codes in a logical basis remains experimentally challenging.

As illustrated in Fig. \ref{figbp}, several noise processes may occur at different stages of the protocol, potentially degrading the quality of the shared entangled pairs in realistic implementations.  In our analysis, we account for the dominant noise sources affecting the protocol, including the light–matter interaction efficiency $\eta_{int}$, the combined fiber-coupling and frequency conversion efficiency $\eta_c$, and the detector efficiency $\eta_d$.  Furthermore, achieving the theoretical upper bound of USD for 1-loss cat codes remains experimentally unclear. For practical implementation, we adopt the linear-optics solution proposed in \cite{Li2023a}, noting that its success probability does not reach the theoretical limit. For the 0-loss cat codes, the theoretical upper bound of USD can be easily achieved  using a beam splitter.

Since the codewords are generally non-orthogonal,  USD cannot be performed deterministically, and the theoretical maximum success probability is given by 
$1-|\braket{\phi|\psi}|$,  where $\ket{\phi}$ and $\ket{\psi}$ denote the states to be distinguished.

\section{\label{sec:bell}Theoretical analysis}

Having established the cavity-QED setup, we now derive the explicit form of the Bell inequality relevant to our protocol.  After performing the USD measurement on the light modes, Alice and Bob share the following entangled atomic state, on which the Bell-CHSH inequality can be tested \cite{Li2023a}:
\begin{eqnarray} \label{eq3}
    \hat{\rho}_{fi}=F\ket{\phi^+}\bra{\phi^+}+(1-F)\ket{\phi^-}\bra{\phi^-},
\end{eqnarray}
where $\ket{\phi^\pm}=(\ket{\uparrow\uparrow}\pm\ket{\downarrow\downarrow})/\sqrt{2}$ are the Bell states, and $F$ denotes the fidelity with respect to the target Bell state $\ket{\phi^+}$. 

In this work, we adopt the widely used CHSH formulation \cite{Clauser1969}, for which the Bell-CHSH inequality reads:
\begin{eqnarray} \label{eq4}
    \begin{aligned}
        S=tr[\rho_{AB}(&\hat{A}_0\otimes\hat{B}_0+\hat{A}_0\otimes\hat{B}_1\\
        &+\hat{A}_1\otimes\hat{B}_0-\hat{A}_1\otimes\hat{B}_1)]\leq2,
    \end{aligned}
\end{eqnarray}
where $\{\hat{A}_0, \hat{A}_1\}$ and $\{\hat{B}_0, \hat{B}_1\}$ denote the measurement bases chosen by Alice and Bob, respectively.   For the mixed state (\ref{eq3}), the CHSH parameter $S$ is maximized with the measurement settings \cite{Horodecki1995}: $\hat{A}_0=\hat{\sigma}_z, \hat{A}_1=\hat{\sigma}_x,\hat{B}_0=\cos{\theta}\hat{\sigma}_z+\sin{\theta}\hat{\sigma}_x, \hat{B}_1=\cos{\theta}\hat{\sigma}_z-\sin{\theta}\hat{\sigma}_x$ with $\theta$ defined by $\tan{\theta}=2F-1$. The corresponding maximal value of $S$ is then
\begin{eqnarray} \label{eq6}
    S_{max}=2\sqrt{1+(2F-1)^2}.
\end{eqnarray}
Hence, when $F>1/2$, the CHSH inequality is violated, meaning $S_{max}>2$.  Such a violation confirms that the correlations between Alice's and Bob's atomic qubits cannot be explained by any local hidden-variable model, demonstrating nonlocality mediated via the cavity-QED atom-light interface rather than direct photon-photon entanglement.

\subsection{\label{dik}DI-QKD secret key rate}

While violations of Bell inequalities are crucial for establishing device-independent security, achieving sufficiently large violations remains experimentally challenging.  The high CHSH values predicted in our scheme indicate that it can meet the requirements for DI-QKD, positioning it as a promising candidate for device-independent quantum communication.

We next evaluate the SKR for a standard DI-QKD protocol using the expression \cite{Acin2007,Pironio2009}:
\begin{eqnarray} \label{eq8}
    R=R_{eg}(1-h(Q)-\chi(S)),
\end{eqnarray}
where $Q$ is the quantum bit error rate (QBER) and $R_{eg}$ denotes the rate of successful entangled state generation.  Assuming Alice and Bob are separated by a distance $L$, and that the repetition rate of the slowest gate or measurement operation in the protocol is $t_0$, the entanglement generation rate can be expressed as 
\begin{eqnarray}
    R_{eg}=P/\max(L/c_f,t_0),
\end{eqnarray}
where $c_f=2\times10^8$ m/s is the speed of light in optical fiber at telecom wavelength.  The function $h(x)=-x\log_{2}{x}-(1-x)\log_{2}{(1-x)}$ is the binary entropy, and $\chi(S)$ quantifies the adversary's information leakage, given by
\begin{eqnarray} \label{eq9}
    \chi(S)=h\left(\frac{1+\sqrt{(S/2)^2-1}}{2}\right).
\end{eqnarray}

In the next section, we use (\ref{eq6}) to quantify the achievable Bell-CHSH violation as a function of key physical parameters, including $\alpha$, distance, memory coherence time, and realistic inefficiencies, and simultaneously compute the SKR under realistic conditions using (\ref{eq8}) to assess the performance of our protocol for DI-QKD applications. This combined analysis allows us to identify the operational regime in which nonlocality can be experimentally observed and the protocol can achieve practical device-independent quantum key distribution.

\section{\label{per}Performance}

Let us now evaluate the performance of our protocol for implementing a Bell test under realistic experimental imperfections and compute the SKR for a standard DI-QKD protocol. In subsection \ref{bn}, we analyze the protocol's behavior under various noise conditions and compare the results obtained for cat codes of different loss orders.  Using (\ref{eq6}) we assess performance by calculating the CHSH parameter $S$. The coherent state amplitude $\alpha$ influences both the fidelity $F$ of the shared atomic state between Alice and Bob and the achievable CHSH value $S$. Further $\alpha$ affects the success probability $P$ of the USD on the cat codewords. 

As discussed in \cite{Li2023a}, a trade-off exists between the fidelity and the entanglement generation rate $R_{eg}$, which in turn induces a trade-off between the CHSH value $S$ and $R_{eg}$ via Eq.~\ref{eq6}. Consequently, Bell violations can, in principle, be observed over arbitrarily long distances if the entanglement generation rate is sufficiently low. To enable a fair comparison across cat codes of different loss orders, we fix $R_{eg}$ and compute $S$ as a function of the separation distance $L$ between Alice and Bob for each loss order.  In subsection \ref{dik}, we then calculate the SKR under realistic conditions to quantify the protocol's performance for DI-QKD applications.

\subsection{\label{bn}Bell inequality violation analysis}
\begin{figure}[hbtp]
   \subfloat[]{
        \includegraphics[width=0.235\textwidth]{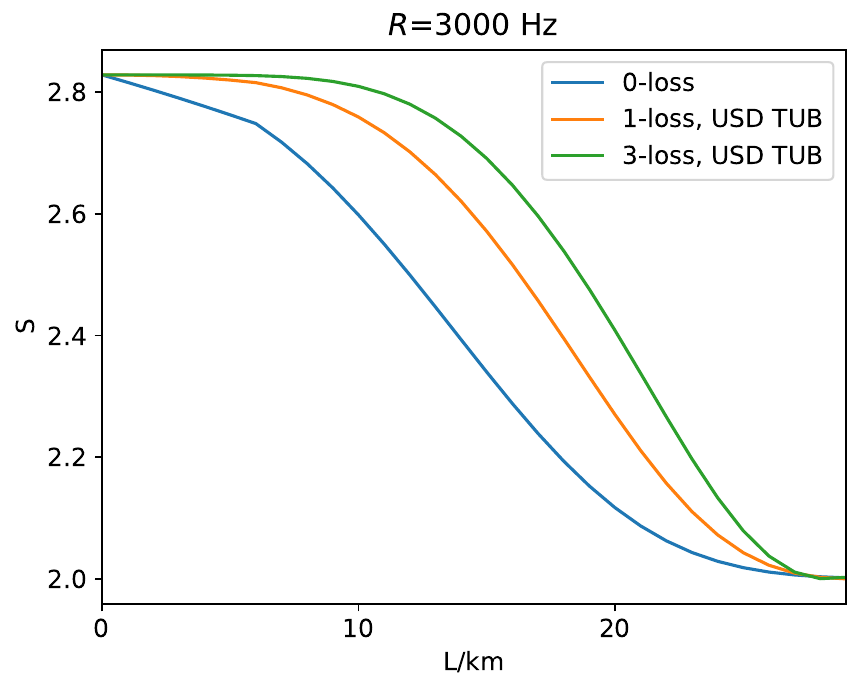}\label{figclo}}
    \subfloat[]{
        \includegraphics[width=0.239\textwidth]{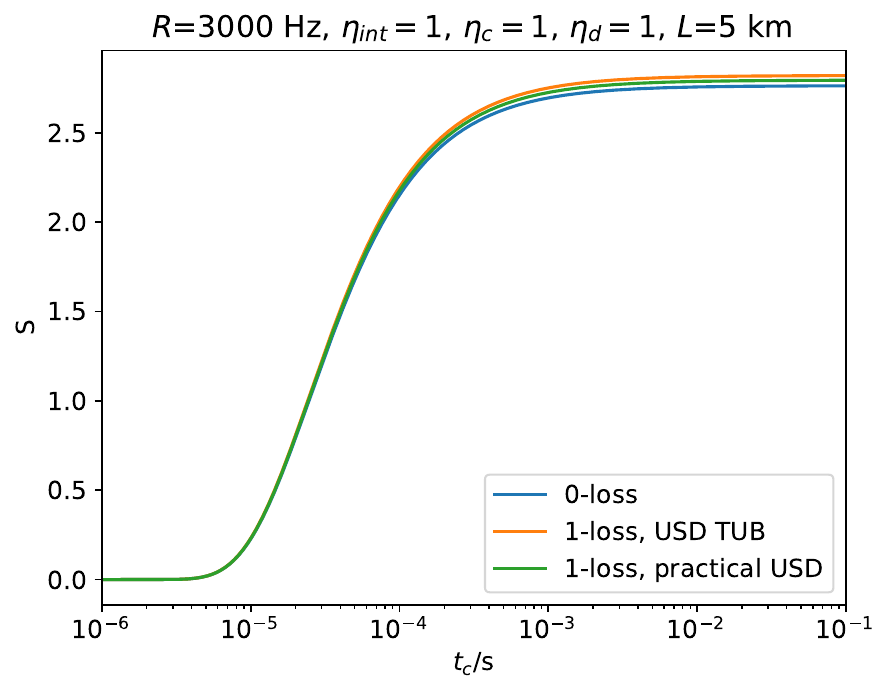}\label{figtc}}
\caption{\label{figclot}   CHSH parameter $S$ under channel-loss-limited conditions. (\textbf{a}) $S$ versus the separation distance $L$ between Alice and Bob. (\textbf{b}) $S$ versus the coherence time $t_c$ of the quantum memory at $L=5$ km. Both panels assume that channel loss is the only error source.}
 \end{figure}

Now let us  compute the value of $S$ for a given entanglement generation rate $R_{eg}$ as a function of the separation distance between Alice and Bob, and compare the performance across different loss orders. Initially, we consider only channel loss errors, assuming all other components are ideal and the USD achieves its theoretical upper bound. The corresponding results under these assumptions are shown in Fig. \ref{figclo}. One can see that at $R=3000$ Hz, the value of $S$ for 1-loss cat codes consistently exceeds that of 0-loss cat codes, and the value of $S$ for 3-loss cat codes likewise surpasses that for 1-loss cat codes. 

Next, we consider the effect of a quantum memory with finite coherence time $t_c$. Fig. \ref{figtc} shows the value of $S$ as a function of $t_c$ at $L=5$ km. As expected, $S$ increases with $t_c$, and the values for 0-loss and 1-loss cat codes are very close at $L=5$ km, in agreement with Fig. \ref{figclo}. Importantly, the results indicate that the CHSH value $S$ saturates for coherence times on the order of $10^{-3}\sim 10^{-2}$ s, suggesting that further increases in $t_c$ offer negligible improvement. Such coherence times are already achievable with state-of-the-art neutral atomic systems \cite{KleineBuening2011}.

\begin{figure*}[htb]
    \subfloat[]{
        \includegraphics[width=0.24\textwidth]{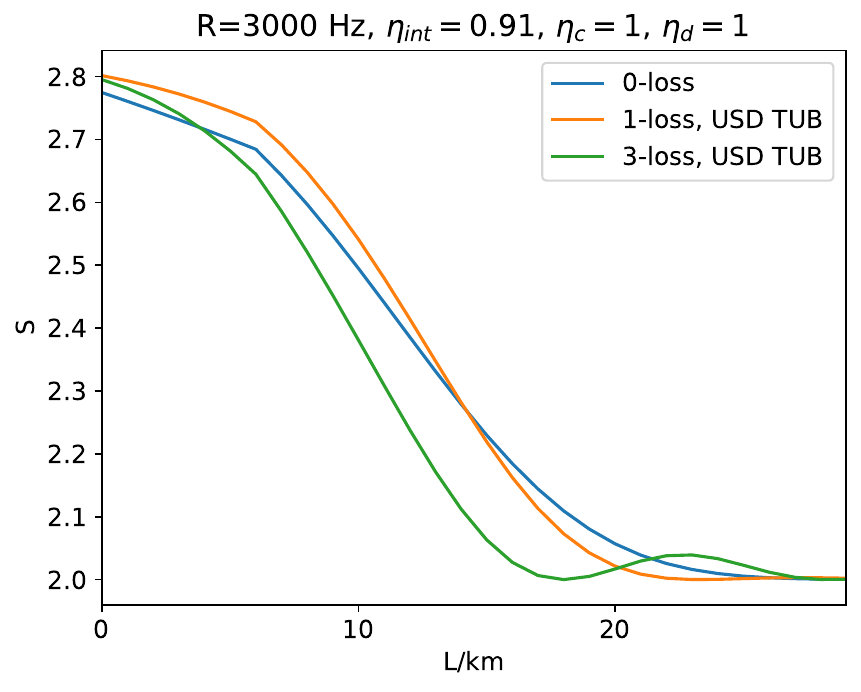}
        \label{dint3}}
    \subfloat[]{
        \includegraphics[width=0.24\textwidth]{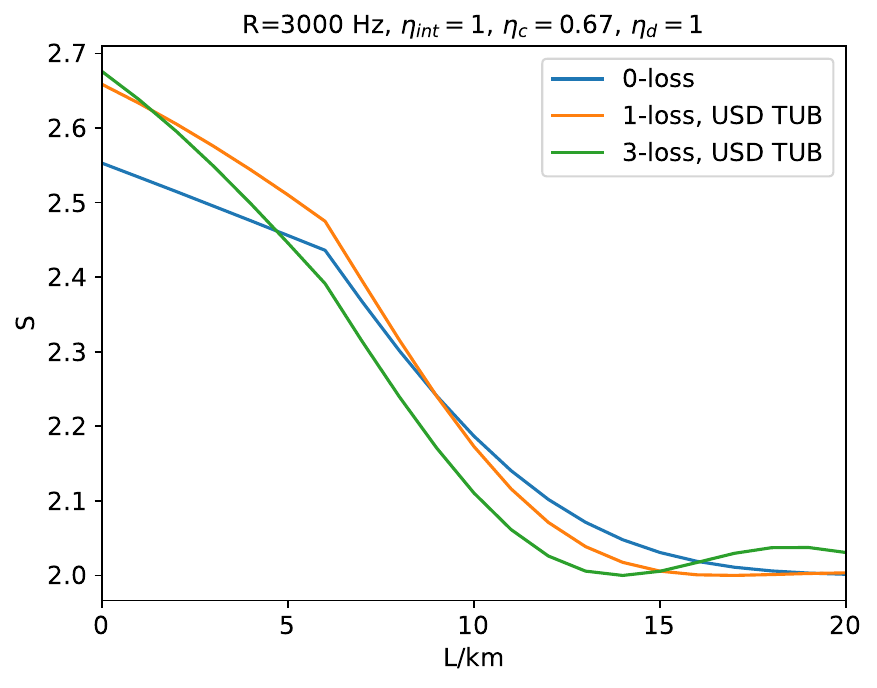}
        \label{dc3}}
    \subfloat[]{
        \includegraphics[width=0.24\textwidth]{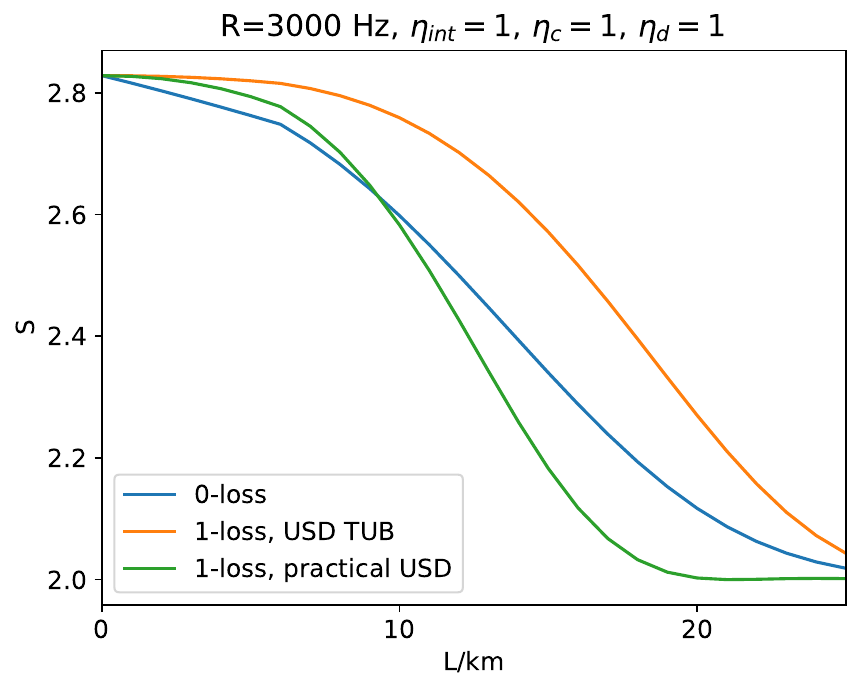}
        \label{ddp}}
    \hfill
    \subfloat[]{
        \includegraphics[width=0.24\textwidth]{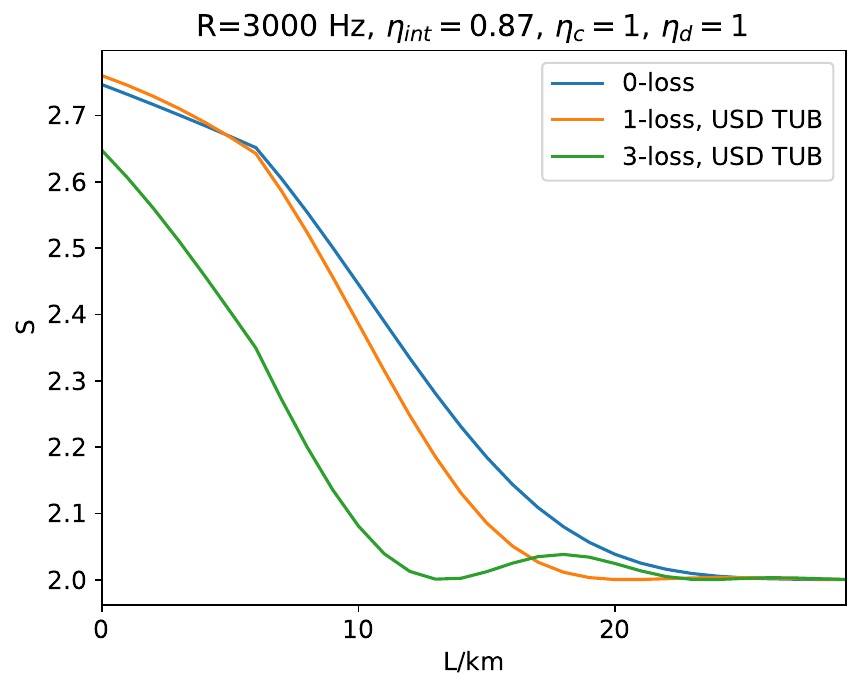}
        \label{dint1}}
    \subfloat[]{
        \includegraphics[width=0.24\textwidth]{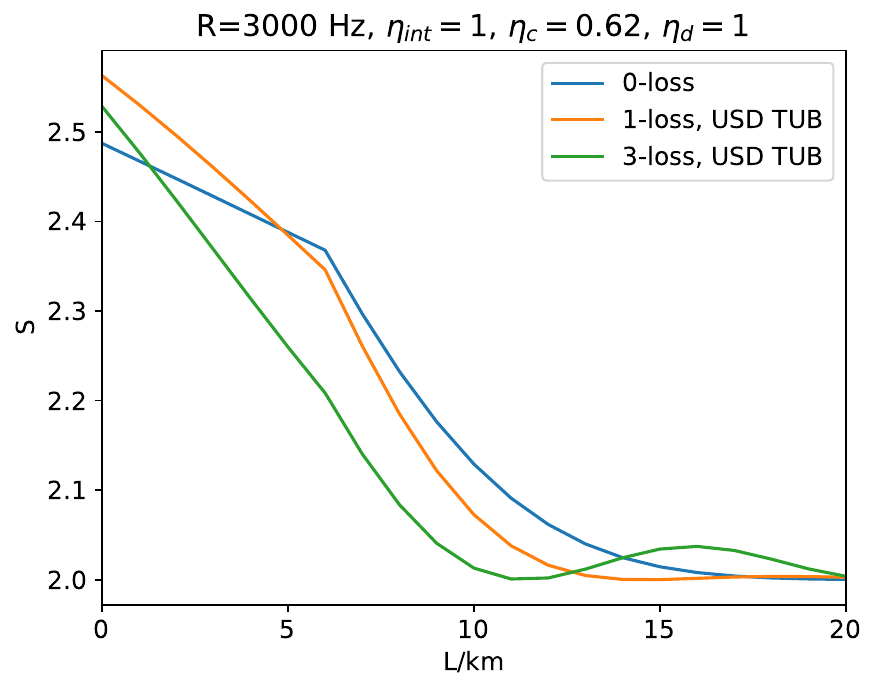}
        \label{dc1}}
    \subfloat[]{
        \includegraphics[width=0.24\textwidth]{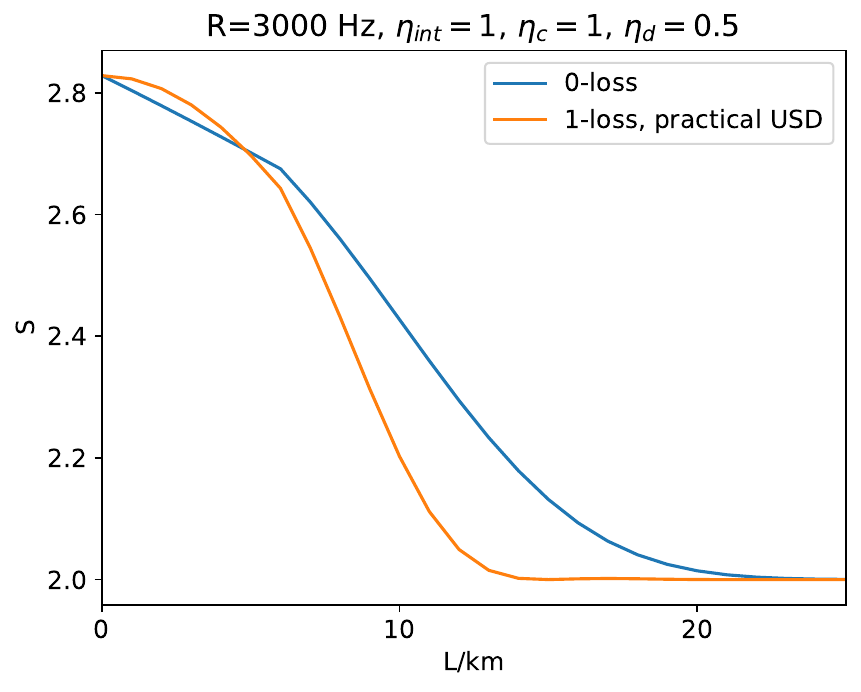}
        \label{dd1}}
     \subfloat[]{
        \includegraphics[width=0.24\textwidth]{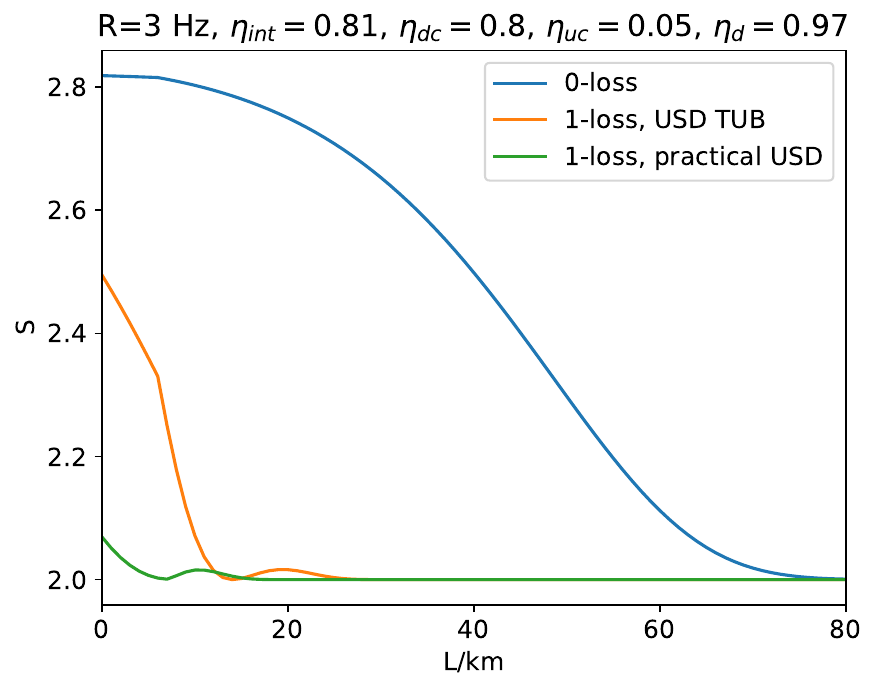}
        \label{figprac}}
\caption{The CHSH value $S$ versus separation distance $L$ for different types of experimental imperfections. In (a) and (d), we consider $\eta_{int}<1$;  in (b) and (e), $\eta_c<1$; 
in (c), we implement a linear-optics solution for the USD with 1-loss cat codes and set $\eta_d=1$;  and in (f), $\eta_d=0.5$.  (g) corresponds to practical parameters in experiment from Table \ref{qep}.  Here, "USD TUB" refers to the theoretical upper bound of the USD success probability. Channel loss is included in all subfigures.}
 \label{figeiec}
\end{figure*}

 Next, we systematically compare the performance of cat codes and coherent-state encoding under realistic experimental inefficiencies, with the goal of identifying the parameter regimes in which cat codes offer a clear advantage. To assess the impact of these imperfections, we vary one error parameter at a time while keeping the others fixed at their ideal values. The results are shown in Fig. \ref{figeiec}. Here, we choose $5$ km as a moderate separation distance, and the comparison between cat codes with different loss orders is mainly restricted to the distance range $L\leq5$ km. An appreciable entanglement generation rate $R=3000$ $\mathrm{Hz}$ is chosen for the performance evaluation. Fig. \ref{dint3} and Fig. \ref{dint1} show the value of $S$ with $\eta_{int}<1$. Compared with Fig. \ref{figclo}, one can see that as $\eta_{int}$ decreases, the 1-loss and 3-loss cat codes gradually lose their advantage over the 0-loss cat codes. For instance, at $\eta_{int}=0.91$, the 3-loss code still outperforms the 0-loss code, while at $L=5$ km, the value of $S$ for 3-loss codes equals that of the 0-loss codes. At $\eta_{int}=0.87$, the 0-loss cat codes outperform 3-loss codes, and the 1-loss codes outperform the 0-loss cat codes but converge at $L=5$ km. Thus, one can expect that 1-loss codes will not outperform the 0-loss cat codes when $\eta_{int}<0.87$. Similarly, Fig. \ref{dc3} and \ref{dc1} show that as $\eta_c$ decreases, the advantages of using 1-loss or 3-loss codes gradually vanish. The critical points at which the 3-loss and 1-loss codes no longer show consistent advantages are $\eta_c=0.67$ and $\eta_c=0.62$, respectively.

Lastly, we consider a physical implementation of the USD for the 0-loss and 1-loss cat codes using linear-optics elements. As shown in Fig. \ref{ddp}, including the linear-optics USD solution for the 1-loss cat codes significantly reduces the performance, even with perfect photon-detection efficiency, yet it still maintains a slight advantage over coherent-state codes for $L<10$ km. When we decrease $\eta_d$, as shown in Fig. \ref{dd1}, the value of $S$ for 1-loss cat codes matches that of the 0-loss cat codes at $\eta_d=0.5$. In conclusion, Fig. \ref{figeiec} clearly identifies the parameter regimes in which employing higher-loss cat codes provides an experimental advantage.

So far, we have identified the error parameters relevant for comparing different codes; however, not all of these values are experimentally feasible with current technology, and they have not yet been simultaneously considered. To analyze the performance of our protocol with current devices, we use experimentally realistic parameter values, as summarized in Table \ref{qep}. Fig. \ref{figprac} shows the value of $S$ versus $L$ at an entanglement distribution rate $R=3$ $\mathrm{Hz}$. Note that in experiment, the achievable efficiencies for up-conversion from telecom to optical $\eta_{uc}$ and down-conversion from optical to telecom $\eta_{dc}$ differ significantly, as shown in Table \ref{qep}.

\begin{table}[hbt]
    \caption{\label{qep}Parameters related to the dominant error sources with the state-of-art technologies.}
    \begin{ruledtabular}
        \begin{tabular}{@{\hskip 10pt}l@{\hskip 10pt}|l@{\hskip 80pt}}
            $\eta_{int}$& 81\% \cite{Hacker2019} \\ \hline
            $\eta_{dc}$ & 80\% \cite{chen2025} \\ 
            \ & 35.6\% \cite{Schaefer2025} \\ \hline
            $\eta_{uc}$& 5\% \cite{Raghunathan_2025} \\ \hline
            $\eta_{d}$& 97\% \cite{Strohauer2025} \\
        \end{tabular}
    \end{ruledtabular}
\end{table}

When all error sources are considered, higher-loss cat codes no longer confer an advantage, whereas coherent-state codes continue to provide robust performance. Notably, a Bell inequality violation ($S>2$) can be achieved at distances up to nearly 80 km. The performance represents a substantial improvement compared to existing Bell test experiments \cite{Hensen2015,Hensen2015,Giustina2015,Rosenfeld2017,Storz2023}, demonstrating the significant potential of employing CV states in Bell test protocols.

\subsection{\label{diqkd}Applications for DI-QKD}

\begin{figure}[t]
    \subfloat[]{
        \includegraphics[width=0.24\textwidth]{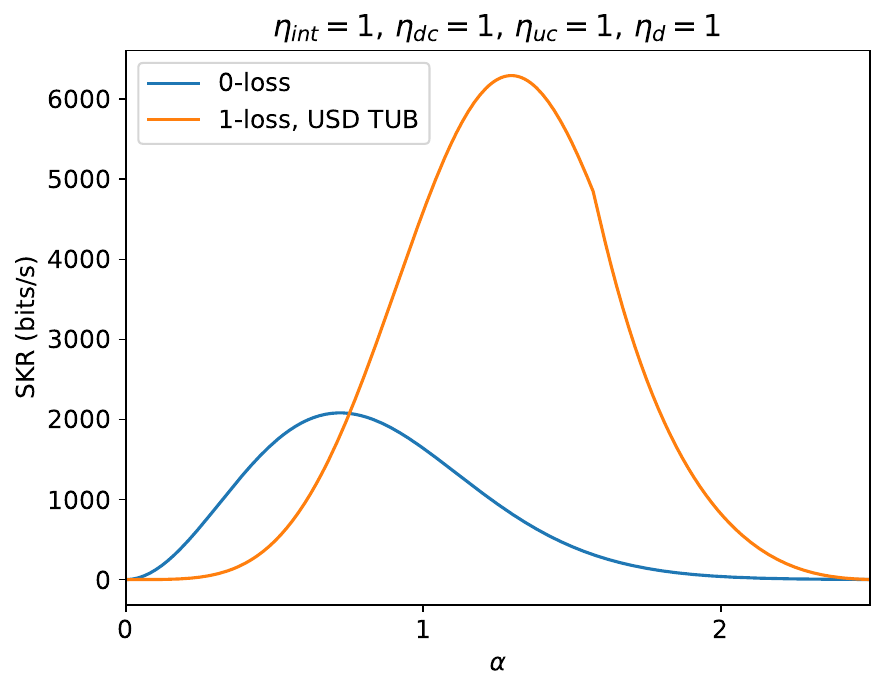}
        \label{qkdid}}
    \subfloat[]{
        \includegraphics[width=0.24\textwidth]{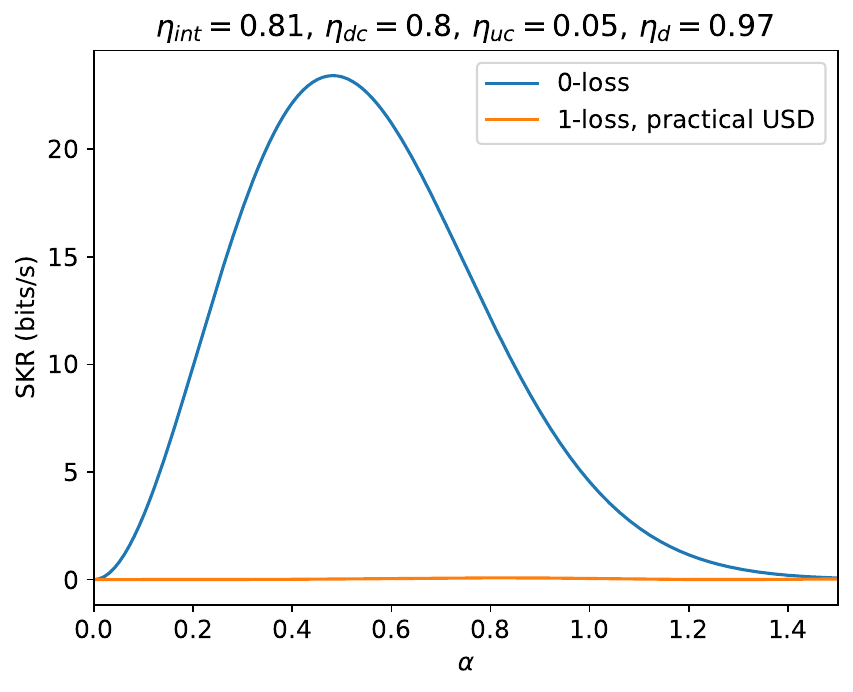}
        \label{qkdlc}}
        \hfill
    \subfloat[]{
        \includegraphics[width=0.24\textwidth]{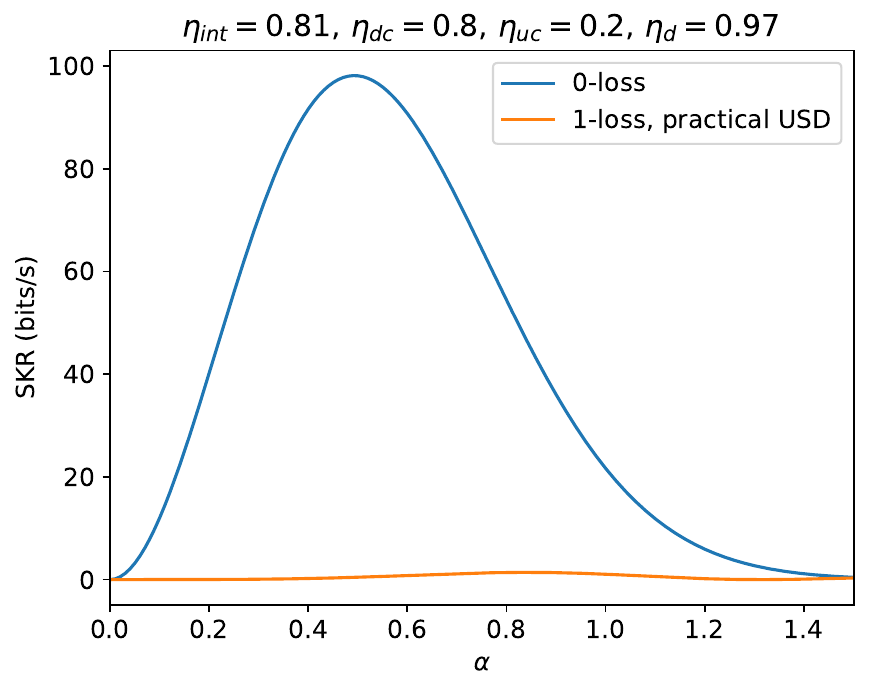}
        \label{qkdhc}}
        \caption{The SKR for DIQKD versus the value of $\alpha$ at $L=10$ km with different error parameters. In (a), only the channel loss is included; in (b), we use the practical noise parameters from Table \ref{qep}; and in (c), we increase the up-conversion efficiency to $\eta_{uc}=0.2$.}
        \label{figdi}
\end{figure}

Fig.~\ref{figdi} presents the SKR for DI-QKD protocols with different noise conditions. 
As shown in Fig.~\ref{qkdid}, the SKR can reach nearly 6000 bits/s for 1-loss cat codes, while it reaches only around 2000 bits/s for 0-loss cat codes.  Thus, 1-loss cat codes outperform 0-loss cat codes when only channel loss and detection inefficiencies are considered.  However, using the practical noise parameters from Table~\ref{qep}, as shown in Fig.~\ref{qkdlc}, the SKR for 1-loss cat codes drops almost to zero, whereas it remains around 25 bits/s for 0-loss cat codes.  This SKR is comparable to values achieved in previous experimental demonstrations \cite{Nadlinger2022,Zhang2022}, despite the longer transmission distance considered here.  Furthermore, if the up-conversion efficiency $\eta_{uc}$ can be improved in future experiments—for instance, reaching $\eta_{uc}=0.2$ as shown in Fig.~\ref{qkdhc}—the SKR can increase to approximately 100 bits/s, surpassing existing protocols.

\section{\label{con}Concluding Discussion}

In this work, we have proposed and thoroughly analyzed a Bell test protocol based on coherent states in cavity-QED systems. 
By accounting for realistic experimental imperfections—such as light-matter interaction losses, fiber-coupling inefficiencies, detector limitations, and non-ideal USD—we have evaluated the performance of our scheme under both idealized and experimentally feasible conditions. 
We also identify the operational thresholds for employing error-correcting cat codes under each type of imperfection. 
Our results demonstrate that this protocol can enable reliable long-distance entanglement distribution and Bell inequality violation, with violations achievable up to nearly 80 km using state-of-the-art devices. 
We further assess the applicability of our protocol for DI-QKD. 
The analysis shows that the achievable secret key rates are comparable to existing implementations and could be further enhanced with improved experimental parameters.

Overall, this study indicates that continuous-variable encodings combined with cavity-QED platforms constitute a promising approach for both foundational tests of quantum mechanics and practical quantum communication. While 1-loss cat codes do not currently offer advantages with existing device parameters, they retain potential benefits for future implementations. For example, employing atoms/ions with telecom transition frequencies could eliminate the need for frequency conversion, and light-matter interaction efficiencies $\eta_{int}$ could be improved through advanced cavity design and fabrication. Future directions may include exploring more general RSBCs, such as binomial codes or squeezed cat codes. Incorporating quantum repeaters could further extend the distribution distance and enhance the quality of entangled pairs. In summary, this work represents an intermediate step toward the practical realization of long-distance quantum communication architectures leveraging coherent states.

\begin{acknowledgments}
We thank Shin Sun and Yuwei Zhu for valuable discussions. This work was supported by the Moonshot R\&D Program Grants JPMJMS2061 \& JPMJMS226C and the JSPS KAKENHI Grants No. 21H04880 and No. 24K07485.
S.B. and H.J. is supported by the National Research Foundation of Korea (NRF) grant funded by the Korea government (MSIT) (Nos. RS-2024-00413957, RS-2024-00438415, and RS-2023-NR076733) and by the Institute of Information \& Communications Technology Planning \& Evaluation (IITP) grants funded by the Korea government (MSIT) (IITP-2025-RS-2020-II201606 and IITP-2025-RS-2024-00437191).
\end{acknowledgments}

\section*{Conflict of Interest}
The authors declare no conflict of interest.

\bibliography{QP2}
\end{document}